\documentclass[twocolumn,showpacs,preprintnumbers,amsmath,amssymb,letterpaper]{revtex4}
\usepackage{graphicx}
\usepackage{dcolumn}
\usepackage{bm}
\usepackage{amssymb}
\usepackage{epsfig}

\newcommand{\beq}{\begin{equation}}
\newcommand{\eeq}{\end{equation}}
\newcommand{\bea}{\begin{eqnarray}}
\newcommand{\eea}{\end{eqnarray}}
\newcommand{\dif}{\mathrm{d}}

\newcommand{\ms}{\overline{\mathrm{MS}}}

\newcommand{\sumint}[1]{\hbox{$\sum$}\!\!\!\!\!\!\!\int_{#1}}
\newcommand{\sumintp}[1]{\hbox{$\sum{ }'$}\!\!\!\!\!\!\!\!\!\!\int_{#1}}

\newcommand{\Lm}{\Lambda}
\newcommand{\Lbar}{\bar{\Lm}}

\newcommand{\PiL}{\overline{\Pi}_L(P)}
\newcommand{\PiT}{\overline{\Pi}_T(P)}
\newcommand{\PiLaT}{\overline{\Pi}_{L/T}(P)}
\newcommand{\PiLT}{\overline{\Pi}_L^{(T)}}
\newcommand{\PiTT}{\overline{\Pi}_T^{(T)}}
\newcommand{\PiLaTT}{\overline{\Pi}_{L/T}^{(T)}}

\newcommand{\PiLTZ}{\overline{\Pi}_{L/T}^{(0)}(P)}
\newcommand{\PiZ}{\overline{\Pi}^{(0)}(P)}

\newcommand{\Ib}{\mathcal{I}}
\newcommand{\If}{\widetilde{\mathcal{I}}}

\begin{document}

\preprint{\hskip5.5in\vbox{CERN-PH-TH-2009-172\\ TUW-09-14}}

\title{The $N_{\rm f}^3 g^6$ term in the pressure of hot QCD}
\author{A.~Gynther$^{1}$}
\email{gynthera@hep.itp.tuwien.ac.at}

\author{A.~Kurkela$^{2}$}
\email{kurkela@phys.ethz.ch}

\author{A.~Vuorinen$^{1,3}$}
\email{aleksi.vuorinen@cern.ch}

\affiliation{$^1$ Institute for Theoretical Physics, TU Vienna, Wiedner Hauptstr. 8-10, A-1040 Vienna, Austria}
\affiliation{$^2$ Institute for Theoretical Physics, ETH Zurich, CH-8093 Zurich, Switzerland}
\affiliation{$^3$ CERN, Physics Department, TH Unit, CH-1211 Geneva 23, Switzerland}

\begin{abstract}
\noindent We determine the first independent part of the $g^6$ coefficient in the weak coupling expansion of the QCD pressure at high temperatures, the one proportional to the maximal power of the number of quark flavors $N_{\rm f}$. In addition to introducing and developing computational methods that can be used in evaluating other parts of the expansion, our calculation provides a result that becomes dominant in the limit of large $N_{\rm f}$ and a fixed effective coupling $g_\mathrm{eff}^2 \equiv g^2N_{\rm f}/2$.
 \end{abstract}
\date{\today}
\maketitle

\section{Introduction}

The single most fundamental quantity characterizing the bulk properties of a system in thermodynamic equilibrium is its partition function, or equivalently the functional dependence of its pressure $p$ on the temperature $T$ (and other parameters such as the chemical potentials). In the case of hot quantum chromodynamics (QCD), this function has been extensively studied both on the lattice \cite{Aoki:2005vt,Cheng:2007jq} and using perturbation theory, with the purpose of obtaining a consistent description of the quantity all the way from the phase transition region ($T\sim T_c$) to asymptotically high temperatures. At present, the state of the art on the perturbative (high $T$) side is order $g^6\ln\,g$ in the strong coupling constant, which has been reached both at zero and finite quark chemical potentials \cite{Kajantie:2002wa,Vuorinen:2003fs}. In addition, perturbative methods have been successfully applied to a plethora of other quantities (for some recent results, see \textit{e.g.~}Refs.~\cite{Chesler:2009yg,Laine:2005ai,Carrington:2008dw} and references therein).

A persistent problem in perturbative finite temperature QCD is the slow convergence of the various weak coupling expansions. For most quantities, the issue can be traced back to the contributions of the soft ($gT$) and ultrasoft ($g^2T$) energy scales, which respectively enter through the three-dimensional effective theories EQCD and MQCD \cite{Braaten:1995jr}. In the case of the pressure, the contributions of these scales to the ${\mathcal O}(g^6)$ term in the expansion have by now been determined \cite{Kajantie:2003ax,DiRenzo:2006nh,Hietanen:2004ew,Hietanen:2006rc}, and to this end, there is clear motivation for evaluating the part corresponding to the hard ($2\pi T$) scales as well \cite{Laine:2003ay}. This requires the evaluation of all four-loop vacuum diagrams in the full four-dimensional theory, a task that has already been completed in the similar, though technically much simpler, case of scalar $\phi^4$ theory \cite{Gynther:2007bw,Andersen:2009ct}.

The ${\mathcal O}(g^6)$ contribution of the QCD pressure can be divided into several gauge invariant parts, proportional to various group theory invariants, most importantly powers of the number of fundamental fermion flavors $N_{\rm f}$. In the present paper, our purpose is to evaluate the first --- and computationally most straightforward --- of these, the one proportional to the maximal power $N_{\rm f}^3$. Our motivation for this is twofold. On one hand, we wish to demonstrate that the computational machinery built for three-loop QCD and four-loop $\phi^4$ theory calculations in Refs.~\cite{Arnold:1994ps,Arnold:1994eb,Gynther:2007bw} can be straightforwardly applied to the four-loop level of QCD as well. In addition, the $N_{\rm f}^3$ term not only represents the first independent piece of the ${\mathcal O}(g^6)$ coefficient, but in fact becomes dominant over all the other contributions in the limit of a large flavor number, where the effective coupling $g_\mathrm{eff}^2\equiv T_F g^2 = N_{\rm f} g^2/2$ is kept fixed while $N_{\rm f}$ is taken to infinity. In this limit, the theory in fact somewhat trivializes, enabling an all orders numerical evaluation of the partition function \cite{Moore:2002md,Ipp:2003zr,Ipp:2003jy} and providing a rough numerical check for our result.

The paper is organized as follows. In the rest of the first Section, we present our notation and explain, how renormalization is performed in our work. After this, we proceed to review the organization of our calculation in Section II, while the bulk of the detailed computations is left to Section III. Our result is finally assembled and discussed in Section IV, where we in addition analyze the convergence of the weak coupling expansion of the pressure and draw our final conclusions.

\subsection{Notation}

We work in $d=4-2\epsilon$ dimensional Euclidean spacetime, using dimensional regularization to regulate both ultraviolet (UV) and infrared (IR) divergences. We use lower case bold letters to denote three-dimensional spatial vectors and capital letters for four-dimensional spacetime vectors $P = (p_0,\bm p)$, so that $P^2 = P_\mu P_\mu , \, p^2 = p_i p_i$. As usual, sum-integrals are defined by
\bea
&&\int_p  \equiv  \Lm^{2\epsilon}\int\frac{\dif^{d-1}p}{(2\pi)^{d-1}} = \left(\frac{e^{\gamma_{\rm E}} \Lbar^2}{4\pi}\right)\int\frac{\dif^{d-1}p}{(2\pi)^{d-1}}, \\
&&\sumint{P/\{P\}}  \equiv  T\sum_{p_0}\int_p, \quad \sumintp{P/\{P\}} \, \equiv \, T\sum_{p_0\neq 0}\int_p,
\eea
where $\Lm$ is the MS renormalization scale and $\Lbar$ the $\ms$ one, and $P/\{P\}$ refer to bosonic/fermionic Matsubara modes, respectively. Bosonic sum-integrals, in which the $p_0=0$ term has been subtracted out, are marked with a prime, and, borrowing notation from Refs.~\cite{Arnold:1994ps,Braaten:1995jr}, some standard sum-integrals are denoted by
\bea
\Ib_n^m & \equiv & \sumint{P}\frac{(p_0)^m}{(P^2)^n}, \\
\If_n^m & \equiv & \sumint{\{P\}}\frac{(p_0)^m}{(P^2)^n}, \\
\Pi(P) & \equiv & \sumint{Q}\frac{1}{Q^2(Q-P)^2}, \\
\Pi_f(P) & \equiv & \sumint{\{Q\}}\frac{1}{Q^2(Q-P)^2}.
\eea

We perform renormalization by consistently using the bare coupling $g_B$ in all of our calculations, and only in the end expressing the result in terms of the physical, renormalized one via the relation $g_B^2~=~Z_{g} g^2$. We need the parameter $Z_{g}$ to two-loop order, to which it reads
\bea
Z_{g} & = & 1 + \frac{g^2}{(4\pi)^2}\frac{\beta_1}{\epsilon} + \frac{g^4}{(4\pi)^4}\left(\frac{\beta_2}{\epsilon^2}+\frac{\beta_2'}{\epsilon}\right) + \dots
\eea
Here, the coefficients $\beta_i$ can for our purposes be deduced from the known exact beta-function of $N_{\rm f}\rightarrow \infty$ QCD,
\bea
\mu\frac{\partial g^2}{\partial\mu} & = & \frac{T_F g^4}{6\pi^2},\;\;\; T_F \equiv \frac{N_{\rm f}}{2},
\eea
giving us
\bea
\label{betas}
\beta_1 & = & \frac{4T_F}{3},\quad
\beta_2 \,= \,\beta_1^2, \quad
\beta_2' \,= \, 0.
\eea
In this limit, the running coupling correspondingly reads
\bea
\frac{1}{g^2(\Lbar)} & = & \frac{1}{g^2(\mu)}\left(1-\frac{T_Fg^2(\mu)}{6\pi^2}\ln\frac{\Lbar}{\mu}\right),
\eea
which this time is exact to all loop orders. It exhibits the well-known Landau singularity, which implies that large-$N_f$ QCD is well-defined only below some energy scale $\Lambda_L$.

Finally, in the course of our calculation, we will often switch to dimensionless integration variables without explicitly saying so. In all of these cases, the integration momenta and coordinates have been scaled by the appropriate power of $2\pi T$.

\begin{widetext}

\section{Organizing the calculation}

The weak coupling expansion of the QCD pressure is known to order $g^6\ln g$ for an arbitrary number of colors and massless fundamental fermion flavors \cite{Kajantie:2002wa,Vuorinen:2003fs}. In the $N_{\rm f}\rightarrow \infty$ limit (with fixed $g_\mathrm{eff}^2$) and at zero chemical potential, the result reads \cite{Zhai:1995ac}
\bea
\label{pressure}
\frac{p-p_0}{d_A} & = & \frac{\pi^2T^4}{45}\left[-\frac{g_\mathrm{eff}^2}{(4\pi)^2}\frac{25}{2} +\frac{g_\mathrm{eff}^3}{(4\pi)^3} \frac{80}{\sqrt{3}}
+\frac{g_\mathrm{eff}^4}{(4\pi)^4}\left(\frac{100}{3}\ln\frac{\Lbar}{4\pi T} + \frac{5}{3} - 88\ln 2 + 20\gamma_{\rm E} + \frac{80}{3}\frac{\zeta'(-1)}{\zeta(-1)} - \frac{40}{3}\frac{\zeta'(-3)}{\zeta(-3)}\right) \right. \nonumber \\
& & \hspace{1.1cm} \left.-\frac{g_\mathrm{eff}^5}{(4\pi)^5}\frac{320}{\sqrt{3}}\left(\ln\frac{\Lbar}{4\pi T} - \frac{1}{2} + 2\ln 2 + \gamma_{\rm E}\right) + \frac{g_\mathrm{eff}^6}
{(4\pi)^6}p_6 + \mathcal{O}(g^7)\right],
\eea
\end{widetext}
in which we wish to compute the unknown coefficient $p_6$ that is nothing but the $N_f^3 g^6$ term in the full theory pressure. Here, we have subtracted out the pressure of a gas of free quarks and gluons, $p_0 = \pi^2 T^4/45\, (d_A + 7N_{\rm c}N_{\rm f}/4)$, where $d_A\equiv N_{\rm c}^2-1$ and which dominates the pressure in the large $N_{\rm f}$ limit (for all $g_\mathrm{eff}^2$).

\begin{figure*}
\includegraphics[width=100mm]{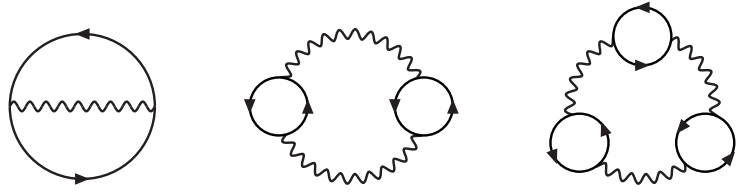}
\caption{The two-, three- and four-loop Feynman diagrams contributing to $p_{\rm E}$ in the large $N_{\rm f}$ limit. The straight lines with an arrow correspond to quarks, and the wavy lines to gluons.}
\label{diags}
\end{figure*}

At leading order in the large $N_{\rm f}$ expansion of $p-p_0$, the effective three-dimensional theories EQCD and MQCD, which incorporate the effects of the soft scales $gT$ and $g^2T$ to the pressure, do not contribute to the coefficients of even powers of $g$. We can thus concentrate on evaluating the contribution of the hard scale, corresponding to the strict (unresummed) perturbation expansion of the pressure in the four-dimensional theory, commonly denoted by $p_{\rm E}$. We note that while this function suffers from IR divergences that for finite $N_{\rm f}$ are canceled by UV divergences in the effective theory contributions, in the large $N_{\rm f}$ limit they are all guaranteed to vanish in dimensional regularization.

In the large $N_{\rm f}$ limit, the function $p_{\rm E}$ has a simple diagrammatic expansion of the form
\bea
\frac{p_{\rm E}}{d_A} & = & \frac{1}{2}Z_{g}g^2I_a + \frac{1}{4}Z_{g}^2g^4I_b + \frac{1}{6}Z_{g}^3g^6I_c + \dots,
\eea
where we have denoted
\bea
I_a & = & -\sumint{P}\frac{\Pi_{\mu\mu}}{P^2}, \\
I_b & = & \sumint{P}\frac{\Pi_{\mu\nu}\Pi_{\nu\mu}}{P^4}, \\
I_c & = & -\sumint{P}\frac{\Pi_{\mu\nu}\Pi_{\nu\rho}\Pi_{\rho\mu}}{P^6},
\eea
and the integrals correspond to the three Feynman graphs in Fig.~\ref{diags}. Here, the one-loop large $N_{\rm f}$ gluon self energy reads
\bea
\Pi_{\mu\nu}(P) & = & -2 T_F\left[2\widetilde{\mathcal{I}}_1^0\delta_{\mu\nu} - (\delta_{\mu\nu}P^2 - P_\mu P_\nu)\Pi_f(P) \right. \nonumber \\
& & \hspace{1cm} \left. - \sumint{\{Q\}}\frac{(2Q-P)_\mu(2Q-P)_\nu}{Q^2(Q-P)^2}\right],
\eea
in which the color indices have been suppressed (the trace over them having already been carried out).

At this point, we may simplify the calculations by dividing the self energy into its three-dimensionally longitudinal and transverse components (for a precise definition, see \textit{e.g.~}Ref.~\cite{Kapusta:2006pm}), in terms of which the above integrals read
\bea
\label{integral_a}
I_a & = & -\sumint{P}\frac{\Pi_L(P)+(d-2)\Pi_T(P)}{P^2}, \\
\label{integral_b}
I_b & = & \sumint{P}\frac{\Pi_L(P)^2+(d-2)\Pi_T(P)^2}{P^4}, \\
\label{integral_c}
I_c & = & -\sumint{P}\frac{\Pi_L(P)^3+(d-2)\Pi_T(P)^3}{P^6}.
\eea
Extracting now the UV divergences of the self-energies by denoting
\bea
\Pi_{L/T}(P) & \equiv & \PiLaT + \frac{\beta_1}{(4\pi)^2}\frac{P^2}{\epsilon}, \label{newpi}
\eea
the expression for the function $p_{\rm E}$ up to order $g^6$ obtains the remarkably simple form
\bea
\frac{p_{\rm E}}{d_A} & = & \frac{1}{2}g^2I_a + \frac{1}{4}g^4\overline{I}_b + \frac{1}{6}g^6\overline{I}_c. \label{pE}
\eea
Here, the bars in $\overline{I}_{b,c}$ denote the replacing of $\Pi_{L/T}(P)$ by $\overline{\Pi}_{L/T}(P)$ (note that with this definition $I_a = \overline{I}_a$), and Eq.~(\ref{betas}) has been used to cancel terms originating from the renormalization corrections against those left over from the redefinition of Eq.~(\ref{newpi}). As $I_a$ and $\overline{I}_b$ are both known, we are only left with the task of computing $\overline{I}_c$, which is furthermore observed to be finite in dimensional regularization.

For notational simplicity, it is convenient to separate the $T=0$ contributions of the self-energies by writing
\bea
\PiLaT & = & \PiLTZ + \PiLaTT(P).
\eea
We find that, in various limits, the two parts of the functions can be written as:
\begin{itemize}
\item at $T=0$:
\bea
\label{Pi0}
\PiLTZ & = & \frac{4T_F}{3}\frac{P^2}{(4\pi)^2}\left[A(\epsilon)\left(\frac{\Lambda^2}{P^2}\right)^\epsilon - \frac{1}{\epsilon}\right] \nonumber \\
& = & \frac{4T_F}{3}\frac{P^2}{(4\pi)^2}\left(\frac{5}{3}+\ln\frac{\Lbar^2}{P^2} + \mathcal{O}(\epsilon)\right), \\
& \equiv & \PiZ \nonumber, \\
A(\epsilon) & = & \frac{6(4\pi)^\epsilon\Gamma(\epsilon)\Gamma(2-\epsilon)^2}{\Gamma(4-2\epsilon)}, \nonumber
\eea
\item at $\epsilon=0$:
\bea
\label{PiLeps0}
\PiLT(P) & = & 2T_F\left(\frac{T^2}{6}\delta_{p_0} +\frac{P^2}{32\pi^3}R^+(n,\tilde{p})\right), \\
\label{PiTeps0}
\PiTT(P) & = &  T_F\frac{P^2}{32\pi^3} R^-(n,\tilde{p}), \\
R^\pm(n,\tilde{p}) & = & \int\frac{d^3r}{r^2}\left(1\pm\frac{1}{r^2}\partial^2_{\tilde{p}}\right)\mathrm{e}^{i\tilde{\bm p}\cdot \bm r}\mathrm{e}^{-|n|r} \nonumber \\
& & \times \left(\mathrm{csch}\, r - \frac{1}{r} + \frac{r}{6}(1-\delta_{n,0})\right) \nonumber, \\
\mathrm{where}\, (n,\,\tilde{\bm p}) & = & (p_0/(2\pi T),\, \bm p/(2\pi T)), \nonumber
\eea
\item at $P\rightarrow\infty$ (up to but not including order $T^6/P^4$):
\bea
\label{Pinfty1}
\PiLT(P) & = & 16T_F\frac{d-2}{d-1}\frac{T^4}{P^2}B(\epsilon), \\
\label{Pinfty2}
\PiTT(P) & = & 16T_F\frac{1}{d-1}\frac{dp_0^2-2P^2}{P^4}T^4\,B(\epsilon), \\
B(\epsilon) & = & \frac{1-2^{2\epsilon-3}}{4\pi^{3/2}}\frac{\Gamma(4-2\epsilon)\zeta(4-2\epsilon)}{\Gamma(3/2-\epsilon)} \left(\frac{4\pi\Lambda^2}{T^2}\right)^\epsilon, \nonumber
\eea
\item at $p\rightarrow 0$ (up to but not including order $p^4$):
\bea
\label{Pzero1}
\PiLT(p_0=0,p) & = & \frac{4T_F}{3}\left\{-3(d-2)\If_1^{\,0} \right. \\
& & \left.+\left[\frac{d-2}{2}\If_2^{\,0} - \frac{A(\epsilon)}{(4\pi)^2}\left(\frac{\Lambda^2}{p^2}\right)^\epsilon\right]p^2\right\}, \nonumber \\
\label{Pzero2}
\PiTT(p_0=0,p) & = & \frac{4T_F}{3}\left[\If_2^{\,0}-\frac{A(\epsilon)}{(4\pi)^2}\left(\frac{\Lambda^2}{p^2}\right)^\epsilon\right]p^2.
\eea
\end{itemize}

The required integral can then be divided into four parts,
\bea
-\overline{I}_c & = & \sumint{P}\frac{\PiL^3 + (d-2)\PiT^3}{P^6} \nonumber \\
& = & \sumint{P}\frac{(d-1)\PiZ^3}{P^6} \nonumber \\
&+& 3\,\sumint{P}\frac{\PiLT(P)+(d-2)\PiTT(P)}{P^6}\PiZ^2 \nonumber \\
& +& 3\,\sumint{P}\frac{\PiLT(P)^2+(d-2)\PiTT(P)^2}{P^6}\PiZ \nonumber \\
& +& \sumint{P}\frac{\PiLT(P)^3+(d-2)\PiTT(P)^3}{P^6}\nonumber \\
& \equiv & K_1 + 3K_2 +3K_3 +K_4. \label{Ic}
\eea
Of these terms, $K_1$ and $K_2$ are UV divergent but IR finite, while $K_3$ and $K_4$ are UV finite but IR divergent. In the following Section, we will set out to evaluate these functions one by one.

\section{Evaluation of the integrals}
\label{results}

In performing the integrals $K_1$--$K_4$, we employ the general strategy of first identifying their UV and IR divergent parts, and then separating them from the rest, writing
\bea
K_i & = & K_i^\mathrm{div} + K_i^\mathrm{fin}.
\eea
The divergent pieces must be evaluated analytically in $d=4-2\epsilon$ dimensions, whereas in the finite integrals we may set $d=4$ and use the integral representations of Eqs.~(\ref{PiLeps0}) and (\ref{PiTeps0}) for the self-energies. Apart from the case of $K_4$, which is treated separately, the subtracted divergent pieces are always identified with specific parts of the integrands of Eqs.~(\ref{PiLeps0}) and (\ref{PiTeps0}), corresponding to terms in the Taylor expansion of the hyperbolic cosecant.

For ease of notation, we write the finite parts of the $K_i$ in terms of dimensionless integrals $M_i$,
\bea
\label{dimfinite}
K_i^\mathrm{fin} & = & \frac{16T_F^3}{9}\frac{T^4}{(4\pi)^4}M_i,
\eea
which one can, when necessary, easily evaluate with numerical tools.

\subsection{$K_1$ integral}

Using the expression of Eq.~(\ref{Pi0}) for $\PiZ$, the integral $K_1$ can be trivially expressed in terms of the $\Ib^m_n$ functions,
\bea
K_1 & = & \frac{T_F^3}{1728\pi^6}(d-1)\left[A(\epsilon)^3\Lambda^{6\epsilon}\Ib_{3\epsilon}^0 - \frac{3}{\epsilon}A(\epsilon)^2\Lambda^{4\epsilon}\Ib_{2\epsilon}^0 \right. \nonumber \\
& & \left. \hspace{2.5cm}+ \frac{3}{\epsilon^2}A(\epsilon)\Lambda^{2\epsilon}\Ib_{\epsilon}^0 \right].
\eea
The values of all of the integrals appearing here are available in the literature (see \textit{e.g.~}Ref.~\cite{Arnold:1994ps}), giving in the end
\bea
K_1&=&\frac{T_F^3T^4}{135(2\pi)^4}\Bigg\{\bigg(\ln\frac{\bar{\Lambda}}{4\pi T}\bigg)^2 + \Big(\frac{13}{3}+2\frac{\zeta'(-3)}{\zeta(-3)}\Big)\ln\frac{\bar{\Lambda}}{4\pi T}\nonumber \\
&+&\frac{209}{36}+\frac{\pi^2}{12}+\frac{13}{3}\frac{\zeta'(-3)}{\zeta(-3)}+\frac{\zeta''(-3)}{\zeta(-3)}\Bigg\}.
\eea

\subsection{$K_2$ integral}

The $K_2$ integral is logarithmically UV divergent due to the integrand behaving like $1/P^4$ at large $P$. Separating thus its large $P$ limit from the rest, we easily find for the divergent part
\bea
K_2^\mathrm{div} & = & 16 T_F \frac{d-2}{d-1}B(\epsilon)\sumint{P}\frac{dp_0^2-P^2}{P^{10}}\PiZ^2 \\
& = & \frac{T_F^3}{9\pi^4}\frac{d-2}{d-1}B(\epsilon) \\
& \times &\!\! \left[d\left(A(\epsilon)^2\Lambda^{4\epsilon}\Ib_{3+2\epsilon}^2 - \frac{2}{\epsilon}A(\epsilon)\Lambda^{2\epsilon}\Ib_{3+\epsilon}^2 + \frac{1}{\epsilon^2}\Ib_3^2\right) \right.\nonumber \\
& & \left.\!\!- \left(A(\epsilon)^2\Lambda^{4\epsilon}\Ib_{2+2\epsilon}^0 - \frac{2}{\epsilon}A(\epsilon)\Lambda^{2\epsilon}\Ib_{2+\epsilon}^0 + \frac{1}{\epsilon^2}\Ib_2^0\right)\right], \nonumber
\eea
where all terms are again readily available.

The finite part of the integral is most conveniently evaluated in two pieces. Starting from the $p_0\neq 0$ terms, we obtain using the notation of Eq.~(\ref{dimfinite})
\bea
\label{M2}
M_2' & = & 2\sum_{n=1}^\infty \int \frac{d^3p}{(2\pi)^3}\left(\frac{5}{3}+\ln\frac{\Lbar^2}{4\pi^2T^2} - \ln(n^2+p^2)\right)^2 \nonumber \\
& & \times \int\frac{d^3r}{r^2}\mathrm{e}^{i\bm p\cdot \bm r} \mathrm{e}^{-nr} \left(\mathrm{csch}\,r-\frac{1}{r}+\frac{r}{6}-\frac{7r^3}{360}\right) \nonumber \\
& \equiv & \left(\frac{5}{3}+\ln\frac{\Lbar^2}{4\pi^2 T^2}\right)^2 M_2'{}^0 \nonumber \\
& & - 2 \left(\frac{5}{3}+\ln\frac{\Lbar^2}{4\pi^2 T^2}\right) M_2'{}^1 + M_2'{}^2, \label{M2parts}
\eea
where we have defined
\bea
\label{M2k}
M_2'{}^{k} & \equiv & 2\sum_{n=1}^\infty \int \frac{d^3p}{(2\pi)^3}\int\frac{d^3r}{r^2}\left[\ln(n^2+p^2)\right]^k\,\mathrm{e}^{i\bm p\cdot \bm r} \mathrm{e}^{-nr} \nonumber \\
& & \times \left(\mathrm{csch}\,r-\frac{1}{r}+\frac{r}{6}-\frac{7r^3}{360}\right).
\eea
Here, the last term inside the parentheses is a result of the UV subtraction we have performed.

Concentrating on the three integrals to be evaluated, we immediately observe that $M_2'{}^0=0$, as the $p$ integral in this case yields $\delta(r)$, while the rest of the integrand vanishes at $r=0$. For $M_2'{}^1$, we on the other hand note that using
\bea
\int \frac{d^3p}{(2\pi)^3} \ln(n^2+p^2)\mathrm{e}^{i\bm p\cdot \bm r} & = & -\frac{\mathrm{e}^{-|n|r}}{2\pi r^3}(1+|n|r),
\eea
both the $p$ integral and the Matsubara sum can be easily performed, leaving us with a straightforwardly solvable hyperbolic integral with the result
\bea
M_2'{}^1 & = & -\frac{1}{120}\left(\frac{43}{9}-9\ln 2 - 7\gamma_{\rm E} + 720\zeta'(-2) \right. \nonumber \\
& & \left.\hspace{1cm} + \frac{50}{3}\frac{\zeta'(-1)}{\zeta(-1)} - \frac{29}{3}\frac{\zeta'(-3)}{\zeta(-3)}\right).
\eea
Finally, the remaining integral $M_2'{}^2$ we evaluate numerically, obtaining
\bea
M_2'{}^2 &=& 0.0099981763.
\eea

Contrary to the $p_0\neq 0$ case, the $p_0=0$ contribution to $K_2$ is UV finite, and the UV subtraction term in fact yields zero in dimensional regularization. The finite dimensionless $M_2^{(p_0=0)}$ is then given by
\bea
\label{M20a}
M_2^{(p_0=0)} & = & \int \frac{d^3p}{(2\pi)^3}\frac{1}{p^2}\left(\frac{5}{3} +\ln\frac{\Lbar^2}{4\pi^2T^2} - \ln p^2\right)^2 \\
&\times &\left[\frac{2\pi}{3}+\int\frac{d^3r}{r^2}\mathrm{e}^{i\bm p\cdot \bm r}p^2\left(\mathrm{csch}\, r - \frac{1}{r}\right)\right], \nonumber
\eea
where we first note that using
\bea
\int\frac{d^3r}{r^2}\mathrm{e}^{i\bm p\cdot \bm r}p^2\frac{r}{6} & = & \frac{2\pi}{3}\left[1+\pi p\delta(p)\right],
\eea
we may write Eq.~(\ref{M20a}) in the form
\bea
\label{M20b}
M_2^{(p_0=0)} & = & \int \frac{d^3p}{(2\pi)^3}\left(\frac{5}{3} +\ln\frac{\Lbar^2}{4\pi^2T^2} - \ln p^2\right)^2 \\
& &\times\int\frac{d^3r}{r^2}\mathrm{e}^{i\bm p\cdot \bm r}\left(\mathrm{csch}\, r - \frac{1}{r} + \frac{r}{6}\right). \nonumber
\eea
We then divide the integral into three parts in a manner analogous to Eq.~(\ref{M2parts}),
\bea
M_2^{(p_0=0)} & = & \left(\frac{5}{3}+\ln\frac{\Lbar^2}{4\pi^2 T^2}\right)^2 M_2^{0,(p_0=0)} \\
&- & 2 \left(\frac{5}{3}+\ln\frac{\Lbar^2}{4\pi^2 T^2}\right) M_2^{1,(p_0=0)} + M_2^{2,(p_0=0)}, \nonumber
\eea
with
\bea
\label{M20k}
M_2^{k,(p_0=0)} & \equiv & \int \frac{d^3p}{(2\pi)^3}\int\frac{d^3r}{r^2}\left[\ln p^2\right]^k\,\mathrm{e}^{i \bm p\cdot \bm r} \nonumber \\
& & \times \left(\mathrm{csch}\,r-\frac{1}{r}+\frac{r}{6}\right).
\eea
The evaluation of these integrals is easily completed,
\bea
M_2^{0,(p_0=0)} & = & 0, \\
M_2^{1,(p_0=0)} & = & 6\zeta'(-2) \\
M_2^{2,(p_0=0)} & = & \frac{2}{\pi^2}\Bigg[\left(-3 + 3\gamma_{\rm E}+\ln\frac{\pi^3}{2}\right)\zeta(3)
- 3\zeta'(3)\Bigg]. \nonumber \\
&&
\eea

Collecting all the different parts of our result, we finally find for $K_2$
\bea
K_2&=&\frac{7T_F^3T^4}{540(2\pi)^4}\Bigg\{\bigg(\ln\frac{\bar{\Lambda}}{4\pi T}\bigg)^2 \nonumber\\
&+&\frac{1}{7} \bigg(\frac{193}{18}-18\ln\,2+\frac{100}{3}\frac{\zeta'(-1)}{\zeta(-1)}\nonumber \\
&-& \frac{58}{3}\frac{\zeta'(-3)}{\zeta(-3)}\bigg)\ln\frac{\bar{\Lambda}}{4\pi T}+14.88558583 \Bigg\}.
\eea

\subsection{$K_3$ integral}

The $K_3$ integral is UV finite, but its $p_0=0$ term contains an IR divergence and must therefore be considered separately. There, the divergent part is easily isolated using Eq.~(\ref{Pzero1}) and is given by
\bea
K_3^\mathrm{div} & = & \frac{64 T_F^3}{3}\left(\If_1^0\right)^2 \frac{T}{(4\pi)^2}(d-2)^2 \\
& & \times \int_p \frac{1}{p^4}\left[A(\epsilon)\left(\frac{\Lambda^2}{p^2}\right)^\epsilon-\frac{1}{\epsilon}\right]. \nonumber
\eea
This integral, however, obviously vanishes in dimensional regularization.

In the IR finite part of the $p_0=0$ term, we can set $\epsilon=0$ and thus get
\bea
K_3^{(p_0=0)} & = & T\!\int\!\frac{d^3p}{(2\pi)^3}\frac{\overline{\Pi}^{(0)}(0,p)}{p^6}\left[\PiLT(0,p)^2 \right. \\
& & \left.\hspace{0.5cm}+2\PiTT(0,p)^2-64T_F^2(\If_1^0)^2\right], \nonumber
\eea
where the last term corresponds to the subtraction of the divergence. This yields
\bea
\label{M30}
M_3^{(p_0=0)} & = & \int\!\frac{d^3p}{(2\pi)^3}\left[\frac{5}{3}+\ln\frac{\Lbar^2}{4\pi^2T^2}-\ln p^2\right] \\
& & \times\left[\frac{3}{16\pi}\left(2R^+(0,p)^2+R^-(0,p)^2\right) \right. \nonumber \\
& & \left. \hspace{0.5cm}+\frac{1}{p^2}R^+(0,p)\right], \nonumber
\eea
from which we obtain
\bea
M_3^{(p_0=0)} & = & \left(\frac{5}{3}+\ln\frac{\Lbar^2}{4\pi^2 T^2}\right)M_3^{0,(p_0=0)} - M_3^{1,(p_0=0)}, \nonumber \\
& &
\eea
with
\begin{widetext}
\bea
\label{M30k}
M_3^{k,(p_0=0)} & \equiv & \pi \int\frac{d^3p}{(2\pi)^3}\left[\ln p^2\right]^k \left\{3\int_0^\infty dr ds\left[3\frac{\sin(pr)}{pr}\frac{\sin(ps)}{ps}+\frac{1}{p^2}\frac{\sin(pr)}{pr}\frac{\partial^2}{\partial s^2}\left(\frac{\sin(ps)}{ps}\right) \right.\right. \nonumber \\
& & \left. +\frac{1}{p^2}\frac{\partial^2}{\partial r^2}\left(\frac{\sin(pr)}{pr}\right)\frac{\sin(ps)}{ps} + \frac{3}{p^4}\frac{\partial^2}{\partial r^2}\left(\frac{\sin(pr)}{pr}\right)\frac{\partial^2}{\partial s^2}\left(\frac{\sin(ps)}{ps}\right)\right] \times \left(\mathrm{csch}\,r - \frac{1}{r}\right)\left(\mathrm{csch}\,s - \frac{1}{s}\right) \nonumber \\
& & \left.+\frac{4}{p^2}\int_0^\infty dr\left[\frac{\sin(pr)}{pr} + \frac{1}{p^2}\frac{\partial^2}{\partial r^2}\left(\frac{\sin(pr)}{pr}\right)\right]\left(\mathrm{csch}\,r - \frac{1}{r}\right)\right\}.
\eea
A straightforward calculation now gives for the two integrals needed
\bea
M_3^{0,(p_0=0)} & = & \int_0^\infty dr\left[\frac{3}{r^2}\left(\mathrm{csch}\,r - \frac{1}{r}\right)^2  + \frac{1}{r}\left(\mathrm{csch}\,r - \frac{1}{r}\right)\right]
 \;=\;  -6\zeta'(-2) - \ln 2,\\
M_3^{1,(p_0=0)} & = &  1.426271836.
\eea

With the IR finite $p_0\neq 0$ part, we write analogously
\bea
M_3' & = & \left(\frac{5}{3}+\ln\frac{\Lbar^2}{4\pi^2 T^2}\right)M_3'{}^0 - M_3'{}^1,
\eea
where now
\bea
M_3'^k & \equiv & 2\pi\sum_{n=1}^\infty \int\frac{d^3p}{(2\pi)^3}\left[\ln (p^2+n^2)\right]^k \left\{3\int_0^\infty dr ds\left[3\frac{\sin(pr)}{pr}\frac{\sin(ps)}{ps}+\frac{1}{p^2}\frac{\sin(pr)}{pr}\frac{\partial^2}{\partial s^2}\left(\frac{\sin(ps)}{ps}\right) \right.\right. \nonumber \\
& & \hspace{3.4cm}\left. +\frac{1}{p^2}\frac{\partial^2}{\partial r^2}\left(\frac{\sin(pr)}{pr}\right)\frac{\sin(ps)}{ps} + \frac{3}{p^4}\frac{\partial^2}{\partial r^2}\left(\frac{\sin(pr)}{pr}\right)\frac{\partial^2}{\partial s^2}\left(\frac{\sin(ps)}{ps}\right)\right] \nonumber \\
& & \left.\hspace{2cm} \times\, \mathrm{e}^{-n(r+s)} \left(\mathrm{csch}\,r - \frac{1}{r} + \frac{r}{6}\right)\left(\mathrm{csch}\,s - \frac{1}{s} + \frac{s}{6}\right)\right\}.
\eea
\end{widetext}
For $M_3'{}^0$, we get after integrating over $p$ and summing over $n$
\bea
M_3'{}^0 & = & 3\int_0^\infty \frac{dr}{r^2}\left(\mathrm{coth}\,r -1\right)\left(\mathrm{csch}\,r -\frac{1}{r} + \frac{r}{6}\right)^2. \nonumber \\
& = & \frac{1}{120}\bigg(-\frac{47}{3}+94\ln 2+10\gamma_{\rm E} - 20\frac{\zeta'(-1)}{\zeta(-1)}\nonumber \\
&+&10\frac{\zeta'(-3)}{\zeta(-3)} + 720\zeta'(-2)\bigg),
\eea
while the remaining integral, $M_3'{}^1$, must again be evaluated numerically with the result
\bea
M_3'{}^1 & = & 0.0006015294.
\eea

Summing up all the different pieces, we finally get for the entire $K_3$ integral
\bea
K_3&=&-\frac{T_F^3T^4}{135(2\pi)^4} \Bigg\{\bigg(\frac{47}{12}+\frac{13}{2}\ln\,2-\frac{5}{2}\gamma_{\rm E}+5\frac{\zeta'(-1)}{\zeta(-1)}\nonumber \\
&-& \frac{5}{2}\frac{\zeta'(-3)}{\zeta(-3)}\bigg)\ln\frac{\bar{\Lambda}}{4\pi T}+44.74417803\Bigg\}.
\eea

\subsection{$K_4$ integral}

With the integral $K_4$, we again have an IR divergence in the $p_0=0$ term, which we attempt to separate first. Using Eq.~(\ref{Pzero1}), we get
\bea
K_4^\mathrm{div} & = & 64T_F^3 T (d-2)^3 \left(\If_1^{\,0}\right)^2 \int_p \frac{1}{p^6} \\
& \times & \left\{-\If_1^{\,0}+\left[\frac{1}{2}\If_2^{\,0}-\frac{A(\epsilon)}{(4\pi)^2(d-2)}\left(\frac{\Lambda^2}{p^2}\right)^\epsilon\right]p^2\right\}, \nonumber
\eea
which, just like $K_3^\mathrm{div}$, vanishes in dimensional regularization.

For the IR finite part of the $p_0=0$ term, we on the other hand obtain after subtracting off all the IR divergent parts
\bea
\label{M40}
M_4^{(p_0=0)} & = & \int\frac{d^3p}{(2\pi)^3}\frac{1}{p^4} \nonumber \\
& & \times\left[\frac{3}{2}R^+(0,p)-4\pi\left(\ln p - \frac{4}{3}+\ln 2 + \gamma_{\rm E}\right) \right. \nonumber \\
& & \left.+\frac{9p^2}{128\pi^2}\Big(16\pi R^+(0,p)^2 + 4p^2R^+(0,p)^3\right. \nonumber \\
& & \left.\hspace{1.5cm}+p^2R^-(0,p)^3\Big)\right].
\eea
Here, we identify the potential problem that as the two first terms inside the square parentheses are separately divergent and only converge when summed together, the numerical evaluation of the integral is inconvenient in its present form. To this end, we consider the first term in more detail.

By carrying out the angular integrals, we find for the function $R^+(0,p)$
\bea
R^+(0,p) & = & 8\pi\int_0^\infty \frac{dr}{(pr)^3}\Big(\sin(pr)-pr\cos(pr)\Big) \nonumber \\
& & \hspace{1.4cm}\times \left(\mathrm{csch}\,r - \frac{1}{r}\right),
\eea
to which we add and subtract a ($p$-independent) term in such a way that its small $p$ behavior becomes apparent. A straightforward calculation now gives
\bea
\label{Rplus}
\frac{R^+(0,p)}{8\pi} & = & \frac{1}{3}\left(\ln p - \frac{4}{3}+\ln\,2+\gamma_{\rm E}\right) \nonumber \\
& + & \int_0^\infty\!\! \frac{dr}{(pr)^3}\left(\sin(pr)-pr\cos(pr)-\frac{1}{3}(pr)^3\right) \nonumber \\
& & \hspace{1.5cm}\times\mathrm{csch}\,r,
\eea
where the first term obviously cancels against the second term of Eq.~(\ref{M40}).

Putting everything together, we obtain
\bea
\!\!\!\!\!\!M_4^{(p_0=0)} & = & M_4^{1,(p_0=0)} + M_4^{2,(p_0=0)} + M_4^{3,(p_0=0)},
\eea
where
\begin{widetext}
\bea
M_4^{1,(p_0=0)} & = & 12\pi\int\frac{d^3p}{(2\pi)^3}\frac{1}{p^4}\int \frac{dr}{(pr)^3}\left(\sin(pr)-pr\cos(pr)-\frac{1}{3}(pr)^3\right)\mathrm{csch}\,r \nonumber \\
& = & -\frac{3\pi^2}{32}, \\
M_4^{2,(p_0=0)} & = & 18\pi\int\frac{d^3p}{(2\pi)^3}\frac{1}{p^2}\int_0^\infty dr\,ds\left[1 + \frac{1}{p^2}\partial_r^2 + \frac{1}{p^2}\partial_s^2 + \frac{1}{p^4}\partial_r^2\partial_s^2\right]\left(\frac{\sin(pr)}{pr}\frac{\sin(ps)}{ps}\right) \nonumber \\
& & \hspace{2.5cm}\times\left(\mathrm{csch}\,r - \frac{1}{r}\right)\left(\mathrm{csch}\,s - \frac{1}{s}\right) \nonumber \\
& = & \frac{3}{5}\int_0^\infty dr\,ds\left[\frac{5r^2-s^2}{r^3}\theta(r-s) + \frac{5s^2-r^2}{r^3}\theta(s-r)\right]\left(\mathrm{csch}\,r - \frac{1}{r}\right)\left(\mathrm{csch}\,s - \frac{1}{s}\right) \nonumber \\ & & \nonumber \\
 & = & 2.938900865, \\
& & \nonumber \\
M_4^{3,(p_0=0)} & = & \frac{9\pi}{2}\int\frac{d^3p}{(2\pi)^3}\int_0^\infty dr\,ds\,dt \nonumber \\
& & \left[5+\frac{3}{p^2}\left(\partial_r^2+\partial_s^2+\partial_t^2\right) + \frac{5}{p^4}\left(\partial_r^2\partial_s^2+\partial_r^2\partial_t^2+\partial_s^2\partial_t^2\right) + \frac{3}{p^6}\partial_r^2\partial_s^2\partial_t^2\right]\left(\frac{\sin(pr)}{pr}\frac{\sin(ps)}{ps}\frac{\sin(pt)}{pt}\right)\nonumber \\
& & \times\left(\mathrm{csch}\,r - \frac{1}{r}\right)\left(\mathrm{csch}\,s - \frac{1}{s}\right)\left(\mathrm{csch}\,t - \frac{1}{t}\right) \nonumber \\
& = & -0.1767993795.
\eea
In all of these formulas, the $r$, $s$ and $t$ derivatives are understood to act only on the $\sin(px)/(px)$ terms.

\begin{figure*}
\includegraphics[width=130mm]{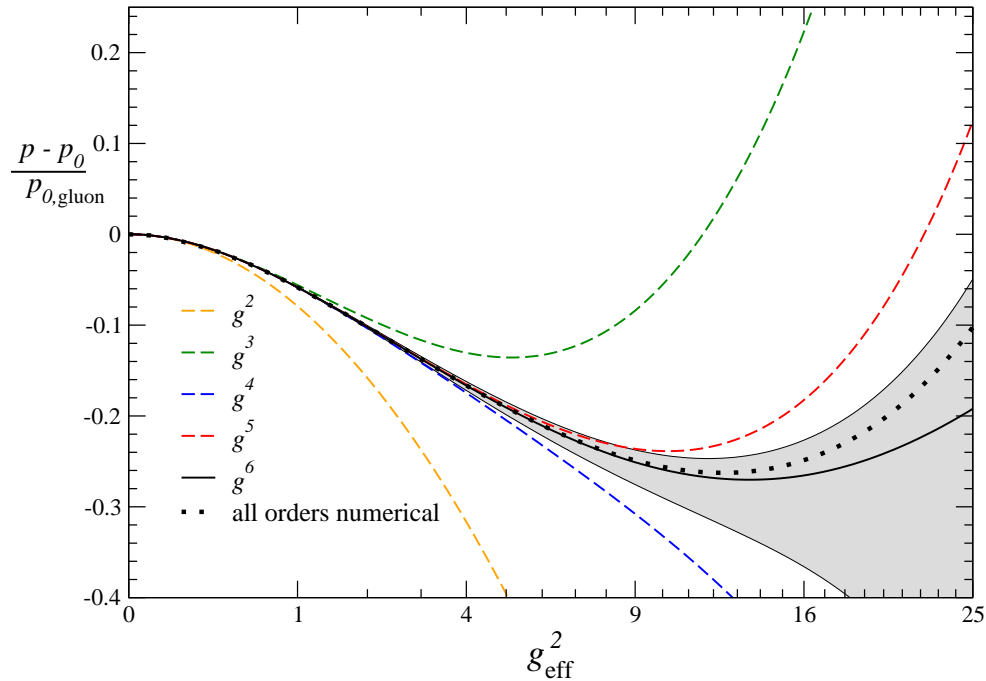}
\caption{[color online] The behavior of the different orders of the weak coupling expansion of the large $N_{\rm f}$ pressure, Eq.~(\ref{pressure}), normalized to the pressure of free gluons $p_{0,\mathrm{gluon}}=d_A\pi^2T^4/45$ and plotted together with the all orders numerical result of Ref.~\cite{Ipp:2003zr}. All the results are evaluated with the renormalization scale $\bar{\Lambda}=e^{-\gamma_{\rm E}}\pi T$, corresponding to the canonical `optimal' scale of Ref.~\cite{Kajantie:1997tt}. For the $g^6$ curve, we also display the effects of varying the renormalization scale by a factor $e^{\gamma_{\rm E}}\approx 1.78$, which corresponds to the shaded grey area. Note that when viewing the pressure as a function of the coupling, also the all orders numerical result has an explicit dependence on the renormalization scale, which only vanishes when taking the running of $g_\mathrm{eff}^2$ into account. This implies that the grey area is not to be compared to the all orders result, but is displayed only to give an idea of the size of the scale dependence of the perturbative result.}
\label{graph}
\end{figure*}

Finally, the $p_0\neq 0$ piece of $K_4$, $M_4'$, is IR finite and in fact closely analogous to $M_4^{3,(p_0=0)}$. It is given by
\bea
M_4' & = & 9\pi \sum_{n=1}^\infty\int\frac{d^3p}{(2\pi)^3}\int_0^\infty dr\,ds\,dt \nonumber \\
& & \left[5+\frac{3}{p^2}\left(\partial_r^2+\partial_s^2+\partial_t^2\right) + \frac{5}{p^4}\left(\partial_r^2\partial_s^2+\partial_r^2\partial_t^2+\partial_s^2\partial_t^2\right) + \frac{3}{p^6}\partial_r^2\partial_s^2\partial_t^2\right]\left(\frac{\sin(pr)}{pr}\frac{\sin(ps)}{ps}\frac{\sin(pt)}{pt}\right) \nonumber \\
& & \times\,\, \mathrm{e}^{-n(r+s+t)}\left(\mathrm{csch}\,r - \frac{1}{r} + \frac{r}{6}\right)\left(\mathrm{csch}\,s - \frac{1}{s}+\frac{s}{6}\right)\left(\mathrm{csch}\,t - \frac{1}{t}+\frac{t}{6}\right) \nonumber \\
& = & 0.00003234178,
\eea
where the $r$, $s$ and $t$ derivatives again act only on the $\sin(px)/(px)$ terms.

Pulling all the pieces together, we find for the entire $K_4$ integral
\bea
K_4&=&\frac{T_F^3T^4}{135(2\pi)^4} \times 27.55287622.
\eea
\end{widetext}

\section{Discussion and conclusions}

In the previous Section, we have evaluated all of the integrals $K_i$ needed in constructing a result for the function $p_6$ of Eq.~(\ref{pressure}). Collecting everything with the help of Eqs.~(\ref{pE}) and (\ref{Ic}), we now finally obtain for the quantity
\bea
p_6 &=& -\frac{800}{9}\ln^2\frac{\Lbar}{4\pi T} - \frac{16}{3}\bigg(\frac{5}{3}+20\gamma_{\rm E}
-88\ln 2 \\
&+& \frac{80}{3}\frac{\zeta'(-1)}{\zeta(-1)}-\frac{40}{3}\frac{\zeta'(-3)}{\zeta(-3)}\bigg)\ln\frac{\Lbar}{4\pi T} + 296.055373. \nonumber
\eea
From here, we may also read off the leading large $N_{\rm f}$ behavior of the function $q_c(N_{\rm f})$ defined in Eq.~(6.11) of Ref.~\cite{Kajantie:2002wa},
\bea
q_c(N_{\rm f}) &=& 0.469861169 \, N_{\rm f}^3+{\mathcal O}(N_{\rm f}^2),
\eea
as well as the coefficient $C_6$ defined in Ref.~\cite{Ipp:2003jy},
\bea
C_6&=&21.8597689.
\eea
It is quite reassuring that our result turned out to lie within the error bars of the very non-trivial numerical estimate $C_6=20(2)$ of Ref.~\cite{Ipp:2003jy}.

The above result can be contrasted on one hand with the exact numerical solution of the large $N_{\rm f}$ pressure \cite{Ipp:2003zr}, and on the other hand to the previous terms in the expansion of Eq.~(\ref{pressure}). In Fig.~\ref{graph}, we perform both comparisons, displaying the perturbative result for the pressure to various orders as a function of the effective coupling $g_\mathrm{eff}$, and comparing it at the same time to the all orders resummation of Ref.~\cite{Ipp:2003zr}. As noted already in previous studies \cite{Ipp:2003zr,Ipp:2003jy}, the convergence of the weak coupling expansion seems quite impressive: The region of applicability of the result clearly increases order by order, and it is only for relatively large values of the effective coupling, $g_\mathrm{eff}^2\gtrsim 20$, that the renormalization scale dependence of the order $g^6$ result becomes strong enough to completely ruin its predictive power.

The good convergence properties of the large $N_{\rm f}$ pressure seem to indicate that in this case, the small parameter in the weak coupling expansion is really $g_\mathrm{eff}^2/(4\pi)^2$, as can indeed be verified from Eq.~(\ref{pressure}). This should be contrasted to the case of full QCD, where the EQCD contributions come with an associated expansion parameter $g$, while the MQCD ones are completely non-perturbative. Remarkably, in the large $N_{\rm f}$ limit, even the effective theory contributions organize themselves in the form of $g_\mathrm{eff}^3$ times a series expansion in $g_\mathrm{eff}^2/(4\pi)^2$. This can be seen to follow from the fact that in the large $N_{\rm f}$ limit, EQCD is a free theory \footnote{This remark is only true for the canonical EQCD action, including interaction terms up to quartic order in the fields and not containing any higher derivative terms. From order $g^7$ onwards, these new terms (which may not be subleading in $1/N_{\rm f}$) start contributing to the pressure.}. Its contribution to the pressure has the form $m_{\rm E}^3\times f(g_3^2/m_{\rm E})$, where $m_{\rm E}\sim g$ is the mass parameter and $g_3^2\sim g^2$ the gauge coupling constant of EQCD, both having weak coupling expansions in powers of $g^2/(4\pi)^2$. In the large $N_{\rm f}$ limit, the ratio $g_3^2/m_{\rm E}$ behaves as $1/N_{\rm f}$, implying that the pressure of EQCD is then indeed merely a number times $m_{\rm E}^3$, and therefore contributes to the pressure of the full theory as $g_\mathrm{eff}^3$ times an expansion in $g_\mathrm{eff}^2/(4\pi)^2$. Finally, the pressure of MQCD is entirely subleading in the large $N_{\rm f}$ limit.

Apart from providing the coefficient of the $N_{\rm f}^3 g^6$ term in the expansion of the QCD pressure, our work has demonstrated the applicability of the computational methods developed for lower order calculations in Refs.~\cite{Arnold:1994ps,Arnold:1994eb} and for the case of scalar $\phi^4$ theory in Ref.~\cite{Gynther:2007bw} in tackling four-loop computations in QCD. We have been able to perform our work semi-analytically, dealing
with all UV and IR divergent parts and logarithms of the renormalization scale analytically in dimensional regularization, and evaluating the remaining finite parts numerically to a high accuracy. At worst, we have had to deal with two-fold numerical integrals, which is a very modest challenge for state of the art computing facilities.

The success of the calculation performed in this paper encourages one to pursue further pieces of the $g^6$ coefficient of the QCD pressure using similar machinery. In particular, the term proportional to $N_{\rm f}^2g^6$ in the result consists of a reasonably limited set of diagrams, a large number of which seem amenable to an analogous treatment. Beyond this, the task rapidly becomes much more complicated, and an approach based on computer algebra methods proves necessary.

\section*{Acknowledgments}

The authors would like to thank Keijo Kajantie, Mikko Laine, Anton Rebhan and York Schr\"oder for useful discussions, and Andreas Ipp both for helpful comments and for providing the data for the all orders numerical result of Ref.~\cite{Ipp:2003zr}. AG was supported by the Austrian Science Foundation FWF, project No.~P19526-N16, AK by the SNF grant 20-122117, and AV~in part by the Austrian Science Foundation, FWF, project No.~M1006, as well as the Sofja Kovalevskaja Award of the Humboldt foundation.

\bibliographystyle{apsrev}
\bibliography{Nf3publ}

\begin{thebibliography}{23}
\expandafter\ifx\csname natexlab\endcsname\relax\def\natexlab#1{#1}\fi
\expandafter\ifx\csname bibnamefont\endcsname\relax
  \def\bibnamefont#1{#1}\fi
\expandafter\ifx\csname bibfnamefont\endcsname\relax
  \def\bibfnamefont#1{#1}\fi
\expandafter\ifx\csname citenamefont\endcsname\relax
  \def\citenamefont#1{#1}\fi
\expandafter\ifx\csname url\endcsname\relax
  \def\url#1{\texttt{#1}}\fi
\expandafter\ifx\csname urlprefix\endcsname\relax\def\urlprefix{URL }\fi
\providecommand{\bibinfo}[2]{#2}
\providecommand{\eprint}[2][]{\url{#2}}

\bibitem[{\citenamefont{Aoki et~al.}(2006)\citenamefont{Aoki, Fodor, Katz, and
  Szabo}}]{Aoki:2005vt}
\bibinfo{author}{\bibfnamefont{Y.}~\bibnamefont{Aoki}},
  \bibinfo{author}{\bibfnamefont{Z.}~\bibnamefont{Fodor}},
  \bibinfo{author}{\bibfnamefont{S.~D.} \bibnamefont{Katz}}, \bibnamefont{and}
  \bibinfo{author}{\bibfnamefont{K.~K.} \bibnamefont{Szabo}},
  \bibinfo{journal}{JHEP} \textbf{\bibinfo{volume}{01}}, \bibinfo{pages}{089}
  (\bibinfo{year}{2006}), \eprint{hep-lat/0510084}.

\bibitem[{\citenamefont{Cheng et~al.}(2008)}]{Cheng:2007jq}
\bibinfo{author}{\bibfnamefont{M.}~\bibnamefont{Cheng}} \bibnamefont{et~al.},
  \bibinfo{journal}{Phys. Rev.} \textbf{\bibinfo{volume}{D77}},
  \bibinfo{pages}{014511} (\bibinfo{year}{2008}), \eprint{0710.0354}.

\bibitem[{\citenamefont{Kajantie
  et~al.}(2003{\natexlab{a}})\citenamefont{Kajantie, Laine, Rummukainen, and
  Schroder}}]{Kajantie:2002wa}
\bibinfo{author}{\bibfnamefont{K.}~\bibnamefont{Kajantie}},
  \bibinfo{author}{\bibfnamefont{M.}~\bibnamefont{Laine}},
  \bibinfo{author}{\bibfnamefont{K.}~\bibnamefont{Rummukainen}},
  \bibnamefont{and} \bibinfo{author}{\bibfnamefont{Y.}~\bibnamefont{Schroder}},
  \bibinfo{journal}{Phys. Rev.} \textbf{\bibinfo{volume}{D67}},
  \bibinfo{pages}{105008} (\bibinfo{year}{2003}{\natexlab{a}}),
  \eprint{hep-ph/0211321}.

\bibitem[{\citenamefont{Vuorinen}(2003)}]{Vuorinen:2003fs}
\bibinfo{author}{\bibfnamefont{A.}~\bibnamefont{Vuorinen}},
  \bibinfo{journal}{Phys. Rev.} \textbf{\bibinfo{volume}{D68}},
  \bibinfo{pages}{054017} (\bibinfo{year}{2003}), \eprint{hep-ph/0305183}.

\bibitem[{\citenamefont{Chesler et~al.}(2009)\citenamefont{Chesler, Gynther,
  and Vuorinen}}]{Chesler:2009yg}
\bibinfo{author}{\bibfnamefont{P.~M.} \bibnamefont{Chesler}},
  \bibinfo{author}{\bibfnamefont{A.}~\bibnamefont{Gynther}}, \bibnamefont{and}
  \bibinfo{author}{\bibfnamefont{A.}~\bibnamefont{Vuorinen}}
  (\bibinfo{year}{2009}), \eprint{0906.3052}.

\bibitem[{\citenamefont{Laine and Schroder}(2005)}]{Laine:2005ai}
\bibinfo{author}{\bibfnamefont{M.}~\bibnamefont{Laine}} \bibnamefont{and}
  \bibinfo{author}{\bibfnamefont{Y.}~\bibnamefont{Schroder}},
  \bibinfo{journal}{JHEP} \textbf{\bibinfo{volume}{03}}, \bibinfo{pages}{067}
  (\bibinfo{year}{2005}), \eprint{hep-ph/0503061}.

\bibitem[{\citenamefont{Carrington et~al.}(2008)\citenamefont{Carrington,
  Gynther, and Pickering}}]{Carrington:2008dw}
\bibinfo{author}{\bibfnamefont{M.~E.} \bibnamefont{Carrington}},
  \bibinfo{author}{\bibfnamefont{A.}~\bibnamefont{Gynther}}, \bibnamefont{and}
  \bibinfo{author}{\bibfnamefont{D.}~\bibnamefont{Pickering}},
  \bibinfo{journal}{Phys. Rev.} \textbf{\bibinfo{volume}{D78}},
  \bibinfo{pages}{045018} (\bibinfo{year}{2008}), \eprint{0805.0170}.

\bibitem[{\citenamefont{Braaten and Nieto}(1996)}]{Braaten:1995jr}
\bibinfo{author}{\bibfnamefont{E.}~\bibnamefont{Braaten}} \bibnamefont{and}
  \bibinfo{author}{\bibfnamefont{A.}~\bibnamefont{Nieto}},
  \bibinfo{journal}{Phys. Rev.} \textbf{\bibinfo{volume}{D53}},
  \bibinfo{pages}{3421} (\bibinfo{year}{1996}), \eprint{hep-ph/9510408}.

\bibitem[{\citenamefont{Kajantie
  et~al.}(2003{\natexlab{b}})\citenamefont{Kajantie, Laine, Rummukainen, and
  Schroder}}]{Kajantie:2003ax}
\bibinfo{author}{\bibfnamefont{K.}~\bibnamefont{Kajantie}},
  \bibinfo{author}{\bibfnamefont{M.}~\bibnamefont{Laine}},
  \bibinfo{author}{\bibfnamefont{K.}~\bibnamefont{Rummukainen}},
  \bibnamefont{and} \bibinfo{author}{\bibfnamefont{Y.}~\bibnamefont{Schroder}},
  \bibinfo{journal}{JHEP} \textbf{\bibinfo{volume}{04}}, \bibinfo{pages}{036}
  (\bibinfo{year}{2003}{\natexlab{b}}), \eprint{hep-ph/0304048}.

\bibitem[{\citenamefont{Di~Renzo et~al.}(2006)\citenamefont{Di~Renzo, Laine,
  Miccio, Schroder, and Torrero}}]{DiRenzo:2006nh}
\bibinfo{author}{\bibfnamefont{F.}~\bibnamefont{Di~Renzo}},
  \bibinfo{author}{\bibfnamefont{M.}~\bibnamefont{Laine}},
  \bibinfo{author}{\bibfnamefont{V.}~\bibnamefont{Miccio}},
  \bibinfo{author}{\bibfnamefont{Y.}~\bibnamefont{Schroder}}, \bibnamefont{and}
  \bibinfo{author}{\bibfnamefont{C.}~\bibnamefont{Torrero}},
  \bibinfo{journal}{JHEP} \textbf{\bibinfo{volume}{07}}, \bibinfo{pages}{026}
  (\bibinfo{year}{2006}), \eprint{hep-ph/0605042}.

\bibitem[{\citenamefont{Hietanen et~al.}(2005)\citenamefont{Hietanen, Kajantie,
  Laine, Rummukainen, and Schroder}}]{Hietanen:2004ew}
\bibinfo{author}{\bibfnamefont{A.}~\bibnamefont{Hietanen}},
  \bibinfo{author}{\bibfnamefont{K.}~\bibnamefont{Kajantie}},
  \bibinfo{author}{\bibfnamefont{M.}~\bibnamefont{Laine}},
  \bibinfo{author}{\bibfnamefont{K.}~\bibnamefont{Rummukainen}},
  \bibnamefont{and} \bibinfo{author}{\bibfnamefont{Y.}~\bibnamefont{Schroder}},
  \bibinfo{journal}{JHEP} \textbf{\bibinfo{volume}{01}}, \bibinfo{pages}{013}
  (\bibinfo{year}{2005}), \eprint{hep-lat/0412008}.

\bibitem[{\citenamefont{Hietanen and Kurkela}(2006)}]{Hietanen:2006rc}
\bibinfo{author}{\bibfnamefont{A.}~\bibnamefont{Hietanen}} \bibnamefont{and}
  \bibinfo{author}{\bibfnamefont{A.}~\bibnamefont{Kurkela}},
  \bibinfo{journal}{JHEP} \textbf{\bibinfo{volume}{11}}, \bibinfo{pages}{060}
  (\bibinfo{year}{2006}), \eprint{hep-lat/0609015}.

\bibitem[{\citenamefont{Laine}(2003)}]{Laine:2003ay}
\bibinfo{author}{\bibfnamefont{M.}~\bibnamefont{Laine}} (\bibinfo{year}{2003}),
  \eprint{hep-ph/0301011}.

\bibitem[{\citenamefont{Gynther et~al.}(2007)\citenamefont{Gynther, Laine,
  Schroder, Torrero, and Vuorinen}}]{Gynther:2007bw}
\bibinfo{author}{\bibfnamefont{A.}~\bibnamefont{Gynther}},
  \bibinfo{author}{\bibfnamefont{M.}~\bibnamefont{Laine}},
  \bibinfo{author}{\bibfnamefont{Y.}~\bibnamefont{Schroder}},
  \bibinfo{author}{\bibfnamefont{C.}~\bibnamefont{Torrero}}, \bibnamefont{and}
  \bibinfo{author}{\bibfnamefont{A.}~\bibnamefont{Vuorinen}},
  \bibinfo{journal}{JHEP} \textbf{\bibinfo{volume}{04}}, \bibinfo{pages}{094}
  (\bibinfo{year}{2007}), \eprint{hep-ph/0703307}.

\bibitem[{\citenamefont{Andersen et~al.}(2009)\citenamefont{Andersen,
  Kyllingstad, and Leganger}}]{Andersen:2009ct}
\bibinfo{author}{\bibfnamefont{J.~O.} \bibnamefont{Andersen}},
  \bibinfo{author}{\bibfnamefont{L.}~\bibnamefont{Kyllingstad}},
  \bibnamefont{and} \bibinfo{author}{\bibfnamefont{L.~E.}
  \bibnamefont{Leganger}} (\bibinfo{year}{2009}), \eprint{0903.4596}.

\bibitem[{\citenamefont{Arnold and Zhai}(1994)}]{Arnold:1994ps}
\bibinfo{author}{\bibfnamefont{P.}~\bibnamefont{Arnold}} \bibnamefont{and}
  \bibinfo{author}{\bibfnamefont{C.-X.} \bibnamefont{Zhai}},
  \bibinfo{journal}{Phys. Rev.} \textbf{\bibinfo{volume}{D50}},
  \bibinfo{pages}{7603} (\bibinfo{year}{1994}), \eprint{hep-ph/9408276}.

\bibitem[{\citenamefont{Arnold and Zhai}(1995)}]{Arnold:1994eb}
\bibinfo{author}{\bibfnamefont{P.}~\bibnamefont{Arnold}} \bibnamefont{and}
  \bibinfo{author}{\bibfnamefont{C.-x.} \bibnamefont{Zhai}},
  \bibinfo{journal}{Phys. Rev.} \textbf{\bibinfo{volume}{D51}},
  \bibinfo{pages}{1906} (\bibinfo{year}{1995}), \eprint{hep-ph/9410360}.

\bibitem[{\citenamefont{Moore}(2002)}]{Moore:2002md}
\bibinfo{author}{\bibfnamefont{G.~D.} \bibnamefont{Moore}},
  \bibinfo{journal}{JHEP} \textbf{\bibinfo{volume}{10}}, \bibinfo{pages}{055}
  (\bibinfo{year}{2002}), \eprint{hep-ph/0209190}.

\bibitem[{\citenamefont{Ipp et~al.}(2003)\citenamefont{Ipp, Moore, and
  Rebhan}}]{Ipp:2003zr}
\bibinfo{author}{\bibfnamefont{A.}~\bibnamefont{Ipp}},
  \bibinfo{author}{\bibfnamefont{G.~D.} \bibnamefont{Moore}}, \bibnamefont{and}
  \bibinfo{author}{\bibfnamefont{A.}~\bibnamefont{Rebhan}},
  \bibinfo{journal}{JHEP} \textbf{\bibinfo{volume}{01}}, \bibinfo{pages}{037}
  (\bibinfo{year}{2003}), \eprint{hep-ph/0301057}.

\bibitem[{\citenamefont{Ipp and Rebhan}(2003)}]{Ipp:2003jy}
\bibinfo{author}{\bibfnamefont{A.}~\bibnamefont{Ipp}} \bibnamefont{and}
  \bibinfo{author}{\bibfnamefont{A.}~\bibnamefont{Rebhan}},
  \bibinfo{journal}{JHEP} \textbf{\bibinfo{volume}{06}}, \bibinfo{pages}{032}
  (\bibinfo{year}{2003}), \eprint{hep-ph/0305030}.

\bibitem[{\citenamefont{Zhai and Kastening}(1995)}]{Zhai:1995ac}
\bibinfo{author}{\bibfnamefont{C.-x.} \bibnamefont{Zhai}} \bibnamefont{and}
  \bibinfo{author}{\bibfnamefont{B.~M.} \bibnamefont{Kastening}},
  \bibinfo{journal}{Phys. Rev.} \textbf{\bibinfo{volume}{D52}},
  \bibinfo{pages}{7232} (\bibinfo{year}{1995}), \eprint{hep-ph/9507380}.

\bibitem[{\citenamefont{Kapusta and Gale}(2006)}]{Kapusta:2006pm}
\bibinfo{author}{\bibfnamefont{J.~I.} \bibnamefont{Kapusta}} \bibnamefont{and}
  \bibinfo{author}{\bibfnamefont{C.}~\bibnamefont{Gale}},
  \emph{\bibinfo{title}{{Finite-temperature field theory: Principles and
  applications}}} (\bibinfo{publisher}{Cambridge, UK: Univ. Pr.},
  \bibinfo{year}{2006}).

\bibitem[{\citenamefont{Kajantie et~al.}(1997)\citenamefont{Kajantie, Laine,
  Rummukainen, and Shaposhnikov}}]{Kajantie:1997tt}
\bibinfo{author}{\bibfnamefont{K.}~\bibnamefont{Kajantie}},
  \bibinfo{author}{\bibfnamefont{M.}~\bibnamefont{Laine}},
  \bibinfo{author}{\bibfnamefont{K.}~\bibnamefont{Rummukainen}},
  \bibnamefont{and} \bibinfo{author}{\bibfnamefont{M.~E.}
  \bibnamefont{Shaposhnikov}}, \bibinfo{journal}{Nucl. Phys.}
  \textbf{\bibinfo{volume}{B503}}, \bibinfo{pages}{357} (\bibinfo{year}{1997}),
  \eprint{hep-ph/9704416}.

\end{thebibliography}

\end{document}